# Dual terahertz frequency combs for photonic RF readout of refractive index sensing with frequency multiplication and active–dummy temperature compensation

Masayuki Higaki[1], Yoshiaki Nakajima[2], Shuji Taue[3], Eiji Hase[4,5], Takeo Minamikawa[4,6], Yu Tokizane[4,5], and Takeshi Yasui[4,5]

[1]Graduate School of Sciences and Technology for Innovation, Tokushima University, 2-1, Minami-Josanjima, Tokushima, Tokushima 770-8506, Japan

[2]Department of Physics, Toho University, 2-2-1 Miyama, Funabashi, Chiba 274-8510, Japan

[3]School of System Engineering, Kochi University of Technology, 185, Miyanokuchi, Tosayamada, Kami, Kochi 782-8502, Japan

[4]Institute of Post-LED Photonics (pLED), Tokushima University, 2-1, Minami-Josanjima, Tokushima, Tokushima 770-8506, Japan

[5]Institute of Photonics and Human Health Frontier (IPHF), Tokushima University, 2-1, Minami-Josanjima, Tokushima, Tokushima 770-8506, Japan

[6]Graduate School of Engineering Science, The University of Osaka, 1-3, Machikaneyama, Toyonaka, Osaka 560-8531, Japan

## Abstract

We present a unified refractive index (RI) sensing platform that integrates THz-comb-based frequency multiplication with dual-comb active–dummy temperature compensation. In conventional RI-sensing optical frequency combs (OFCs), sensitivity, stability, and measurement speed are fundamentally coupled, limiting overall performance. In the proposed system, RI-induced shifts in the repetition frequency are amplified in the terahertz domain, while temperature-induced fluctuations are suppressed through common-mode rejection in a dual-comb configuration. Experimental results demonstrate a sensitivity of $5.05 \times 10^7$ Hz/RIU, high linearity ($R^2$ = 0.9979), improved resolution ($1.07 \times 10^{-4}$ RIU), and high accuracy ($5.50 \times 10^{-5}$ RIU). The RI-induced frequency shift is expanded from tens of hertz to hundreds of kilohertz, enabling rapid and precise readout with short gate times. This approach overcomes the conventional trade-off between sensitivity and stability. More fundamentally, it establishes orthogonal control of signal scaling and noise suppression as a design principle for high-performance RI sensing.

## 1. Introduction

Refractive index (RI) sensing [1] is a fundamental measurement technique that provides sensitive insights into material composition, molecular interactions, and the physical state of a medium, and therefore plays an important role in material characterization [2], chemical analysis [3], and sensing applications such as biological diagnostics [4] and environmental monitoring [5]. In recent years, emerging applications such as real-time biochemical reactions [6], gas sensing [7], and dynamic fluid processes [8] have increasingly demanded RI sensors capable of simultaneously achieving high sensitivity, fast response, and long-term stability. However, achieving these requirements simultaneously remains a significant challenge for conventional optical sensing technologies. Common optical RI sensing techniques include surface plasmon resonance (SPR) sensors [9,10] and fiber-based sensors [11-15]. SPR sensors detect RI changes through shifts in resonance angle or wavelength induced by plasmon excitation at a metal–dielectric interface, offering high sensitivity and broad applicability such as biosensing. Fiber-based sensors rely on RI-dependent changes in guided light propagation in optical fibers, offering advantages such as compactness, robustness, and suitability for remote sensing. Despite these strengths, both approaches fundamentally rely on measuring RI-dependent optical spectral shifts, which ultimately limits their sensing performance in terms of resolution, stability, and speed. This constraint becomes particularly critical when high precision and high-speed measurements are required simultaneously. A promising strategy to overcome this limitation is to decouple signal readout from optical sensing, enabling RI sensing to be performed optically while the signal is retrieved in the radio-frequency (RF) domain. In this domain, mature RF frequency standards and RF instrumentation enable precise, rapid, and versatile measurements.

To enable optical sensing with radio-frequency (RF) readout, optical frequency combs (OFCs) [16-18] have emerged as a promising approach, as they provide a coherent link between optical and electrical frequencies. As a result, OFC-based sensing platforms have become a powerful tool for high-precision RI measurement [19-23] and biosensing [24]. In such platforms, an OFC functions not only as a broadband light source but also as a coherent optical-to-RF frequency converter. For RI sensing, a distinctive feature is the incorporation of a fiber-based RI sensor, such as a multimode interference (MMI) sensor [13-15], into the OFC cavity. Changes in the sample RI induce a shift in the optical spectrum of the OFC via the intracavity MMI sensor. This shift is converted into a change in the effective optical cavity length of the OFC through fiber dispersion. As the repetition frequency ($f_{rep}$) is inversely proportional to the effective optical cavity length, this results in a corresponding change in $f_{rep}$, typically in the tens-of-megahertz range. This $f_{rep}$-encoded scheme, referred to as an RI-sensing OFC, enables direct RF-domain readout of the sensing signal while preserving optical sensing via the intracavity MMI sensor, thereby enabling high-resolution and traceable measurements. Furthermore, placing the sensor inside the cavity enables repeated interaction between the optical field and the sample, effectively increasing the optical path length and thereby improving sensitivity. Nevertheless, two fundamental challenges remain. First, the RI-induced shift of $f_{rep}$ is typically on the order of only tens of hertz per $10^{-3}$ refractive index unit (RIU), which limits frequency resolution—and thus RI resolution—especially at short gate times.

Second, residual temperature fluctuations of the cavity introduce significant drift in the sensor signal (e.g., on the order of hundreds of hertz per degree Celsius), resulting in a time-varying background that degrades measurement reproducibility and increases uncertainty.

To address these challenges, we have developed two complementary strategies that independently target each limitation. First, to overcome the limited sensitivity and slow measurement speed, we exploit frequency multiplication using terahertz (THz) frequency combs [25-27]. By converting the $f_{rep}$ of an RI-sensing OFC into higher-order harmonics in the THz domain, the RI-dependent frequency shift is amplified by factors of thousands, enabling rapid and high-resolution readout beyond the limitations of conventional RF frequency counters. Second, to suppress temperature-induced drift, we employ a dual-comb configuration based on a mechanically shared architecture [28-29], commonly referred to as an active–dummy temperature compensation scheme [22,24]. In this scheme, the active OFC with $f_{rep1}$ is exposed to both sample RI variations and temperature fluctuations, whereas the dummy OFC with $f_{rep2}$ experiences only temperature fluctuations. By monitoring the differential repetition frequency ($\Delta f_{rep} = f_{rep1} - f_{rep2}$) between the active and dummy OFCs, temperature-induced fluctuations are suppressed as common-mode noise, thereby improving measurement reproducibility and reducing uncertainty. Despite their individual success, these approaches still make it difficult to simultaneously achieve high sensitivity, high speed, and temperature stability within a single sensing platform. In particular, frequency multiplication using THz combs amplifies not only the RI-dependent signal but also temperature-induced fluctuations.

In this study, we present a unified RI sensing platform that integrates these two previously independent approaches. Specifically, we combine THz comb–based frequency multiplication [23] with an active–dummy temperature-compensated dual OFC configuration [22,24]. In the proposed system, the RI-dependent shift in $f_{rep}$ is amplified in the THz domain via frequency multiplication, while temperature-induced drift in $f_{rep}$ is suppressed through multi-heterodyne detection using multiple THz harmonics generated by the active and dummy OFCs. This integration decouples temperature stability from sensitivity enhancement, addressing a fundamental limitation of conventional RI-sensing OFCs. As a result, the proposed approach simultaneously achieves robust immunity to temperature drift, ultra-high sensitivity, and millisecond-scale measurement speed. By overcoming the long-standing trade-off between stability and sensitivity, this work establishes a new paradigm for high-performance RI sensing, enabling rapid, precise, and stable detection of dynamic RI changes. From a broader perspective, this work establishes orthogonal control of signal scaling and noise suppression as a fundamental design principle for high-performance RI sensing.

## 2. Principle of operation

In this section, we trace the evolution of the measurement principle leading to the present work, starting from a RI-sensing OFC system and extending to sensitivity enhancement via THz-comb-based frequency multiplication and active–dummy temperature compensation using a dual OFC configuration. Rather than providing a full theoretical derivation given elsewhere [19-24], we focus on the conceptual operation and experimental configuration relevant to the proposed system. By tracing these developments, the technical distinctions and novelty of the proposed approach are clarified.

**RI-sensing OFC.** Figure 1(a) illustrates the operating principle of a RI-sensing OFC [19-21]. The MMI fiber sensor, composed of a clad-less multimode fiber sandwiched between single-mode fibers, acts as an RI-dependent tunable optical bandpass filter. Changes in the sample RI shift the transmission wavelength of the MMI sensor through multimode interference. When incorporated into the OFC cavity as an intracavity element, this spectral shift modifies the oscillation spectrum of the OFC. Owing to fiber dispersion, the spectral shift is translated into a change in the effective optical cavity length ($nL$). Since the repetition frequency is given by $f_{rep}$ = c/(nL), where $c$ is the speed of light in vacuum, this results in a corresponding shift in $f_{rep}$. Thus, RI sensing is achieved via RF-domain readout of $f_{rep}$.

**Sensitivity enhancement via THz-comb-based frequency multiplication.** Figure 1(b) illustrates the operating principle of THz-comb-based frequency multiplication [23]. When the RI-sensing OFC is injected into a photoconductive antenna (PCA) operating in the THz detection (photoconductive mixing) mode, a photoconductive THz (PC-THz) comb with harmonics of the repetition frequency ($f_{rep}$, $2f_{rep}$, •••, $mf_{rep}$) is generated within the PCA. A shift in $f_{rep}$ to $f_{rep} + \delta f_{rep}$ leads to an m-fold amplified frequency shift ($m\delta f_{rep}$) at the $m$-th harmonic. By heterodyning this PC-THz comb with a continuous-wave THz signal of known frequency ($f_{THz}$), the amplified frequency change is converted into an RF beat signal $f_b = |mf_{rep} - f_{THz}|$, whose variation is $\delta f_b = m\delta f_{rep}$, enabling high-sensitivity readout of RI-induced variations.

**Active–dummy temperature compensation using a dual OFC configuration.** Figure 1(c) illustrates the operating principle of active–dummy temperature compensation in a dual OFC configuration [22,24]. Two OFCs, referred to as the active and dummy OFCs, are constructed in a mechanically and thermally shared environment, such that environmental perturbations affect both OFCs in a common-mode manner. The repetition frequency of the active OFC ($f_{rep1}$) responds to both RI variations and temperature fluctuations due to the presence of an intracavity RI sensor, whereas that of the dummy comb ($f_{rep2}$) responds only to temperature variations. By taking the differential repetition frequency ($\Delta f_{rep}$ = $f_{rep1}$ - $f_{rep2}$), temperature-induced common-mode fluctuations are canceled, enabling stable RI sensing.

**Dual-THz-comb-based sensitivity enhancement with active–dummy temperature compensation.** Figure 1(d) illustrates the operating principle of the unified system that combines THz-comb-based frequency multiplication with active–dummy dual OFC-based temperature compensation. Two RI-sensing OFCs, referred to as the active and dummy OFCs, are prepared and injected into two PCAs. As a result, dual PC-THz combs with harmonics of the repetition frequencies ($f_{rep1}$, $2f_{rep1}$, •••, $mf_{rep1}$; $f_{rep2}$, $2f_{rep2}$, •••, $mf_{rep2}$) are generated. A common continuous-wave THz signal with frequency $f_{THz}$ is simultaneously injected into both PCAs, producing dual RF beat signals $f_{beat1}$ and $f_{beat2}$ via heterodyne mixing. After independently measuring the two beat signals ($f_{beat1}$ and $f_{beat2}$) with a RF frequency counter, their differential beat frequency is numerically calculated as $\Delta f_{beat}$ = $f_{beat1}$ - $f_{beat2}$. The variation of this differential beat frequency, $\delta(\Delta f_{beat})$ = $\delta f_{beat}1$ - $\delta f_{beat2}$, corresponds to $m(\delta f_{rep1}$ - $\delta f_{rep2})$. Here, the factor $m$ represents the frequency multiplication, while ($\delta f_{rep1}$ - $\delta f_{rep2}$) corresponds to the temperature-compensated differential signal obtained via the active–dummy scheme. In other words, this approach enables simultaneous sensitivity enhancement and temperature-drift suppression through numerical differential processing in the RF domain.

## 3. Experimental setup

The system consists of dual RI-sensing OFC sources and a photoconductive THz heterodyne detection system, including PCAs and a continuous-wave THz source.

**Dual RI-sensing OFC sources.** Figure 2 shows the configuration of the dual RI-sensing OFC sources used in this study. A pair of MMI fiber sensors is employed as intracavity RI sensing elements in the active and dummy OFCs. The MMI fiber consists of a clad-less multimode fiber (CL-MMF: core diameter = 125 μm, length = 58.94 mm) spliced between two polarization-maintaining single-mode fibers (PMFs: core diameter = 8.5 μm). Based on fourth-order multimode interference, the MMI section acts as an RI-dependent tunable optical bandpass filter centered at the MMI wavelength $\lambda_{MMI}$ = 1556.6 nm, whose transmission wavelength shifts with changes in the surrounding RI. The MMI sensor in the active OFC is placed in a sample cell filled with an aqueous ethanol solution, whereas the MMI sensor in the dummy OFC is placed in a reference medium of pure water. Despite being contained in separate cells, the sample and reference solutions exhibited nearly identical temperature fluctuations.

The linear fiber cavity for the RI-sensing OFC consists of polarization-maintaining single-mode fiber (PMF, PM1550-XP, Thorlabs: dispersion at 1550 nm = 18 ps·km$^{-1}$·nm$^{-1}$), erbium-doped PMF (EDF, PM-ESF-7/125, Coherent: dispersion at 1550 nm =

16 ps·km$^{-1}$·nm$^{-1}$), a wavelength-division multiplexing coupler (WDM, PMWDM-1-9801550-2-B-Q-6, AFR), an optical isolator (ISO, IO-G-1550, Thorlabs), a 90:10 polarization-maintaining output coupler (OC, PMOFM-55-2-B-Q-F-90, AFR), and an intracavity MMI fiber sensor. The EDF is optically pumped by a laser diode (LD, BL976-PAG700, Thorlabs:; wavelength = 976 nm, output power = 700 mW), generating amplified spontaneous emission. Passive mode-locking is achieved using a saturable absorber mirror (SAM, SAM-1550-33-2ps-FC/PC-PMF1550, BATOP: high-reflection band = 1520–1580 nm, absorbance = 33%, modulation depth = 15%, relaxation time constant ≈ 2 ps). The optical cavity lengths of the active and dummy OFCs are adjusted to yield similar repetition frequencies ($f_{rep1}$ and $f_{rep2}$) of approximately 30.69 MHz, with a frequency difference of $\Delta f_{rep}$ = 310 Hz.

The two RI-sensing OFC cavities are mounted on a common copper plate and enclosed in a plastic housing. The copper plate is temperature-stabilized at 20.0 °C using a Peltier heater (TEC1-12708, Kaito Denshi), a thermistor (NXFT15WF104FA2B050, Murata), and a PID-controlled temperature controller (TED200C, Thorlabs), which are not shown in Fig. 2. All components of the dual RI-sensing OFC cavities are arranged to share the same thermal environment, except for the intracavity MMI fiber sensors, thereby ensuring that both OFCs experience nearly identical temperature drifts.

**Photoconductive THz heterodyne detection system.** Figure 3 illustrates the experimental setup of the photoconductive THz heterodyne detection system using dual RI-sensing OFCs. The system consists of three main parts: generation of dual

PC-THz combs, photoconductive THz heterodyne detection with a continuous-wave THz signal, and RF-domain signal processing and measurement.

In the generation of dual PC-THz combs, the optical outputs from the dual RI-sensing OFC sources are individually amplified to several tens of milliwatts using home-built erbium-doped fiber amplifiers (EDFAs). The amplified OFC light is then converted into second-harmonic generation (SHG) light using nonlinear optical crystals ($\beta$-$BaB_2O_4$, thickness = 8 mm), yielding SHG powers of approximately 10 mW at a central wavelength of ~780 nm. After passing through a near-infrared cut filter, the SHG light is focused onto a bowtie-shaped low-temperature-grown GaAs photoconductive antenna (PCA) using an objective lens (magnification = 10, numerical aperture = 0.25). As a result, dual PC-THz combs with frequency spacings corresponding to the repetition frequencies ($f_{rep1}$ or $f_{rep2}$) are generated in the PCAs.

In the photoconductive THz heterodyne detection stage, a narrow-linewidth continuous-wave THz (CW-THz) signal is generated by a frequency multiplier chain (AMC-10-RFH00, Millitech: multiplication factor = 6, output frequency = 101.270 GHz, output power = ~5.0 mW) driven by a microwave synthesizer (E8257D, Keysight Technologies: output frequency = 16.8784 GHz) synchronized with a rubidium (Rb) frequency standard (FS725, Stanford Research Systems: accuracy = $5 \times 10^{-11}$, instability = $2 \times 10^{-11}$ at 1 s). The generated CW-THz wave is collimated and focused onto the PCA using a pair of Teflon lenses (L, LAT100, Thorlabs: diameter = 50 mm, focal length = 91 mm). Through photoconductive mixing between the CW-THz wave and the PC-THz combs, a series of RF beat signals is generated. The lowest-frequency beat component corresponds to the mixing between the CW-THz signal and the nearest harmonic of the repetition frequency. The lowest-frequency beat signals from the two PCAs ($f_{beat1}$ and $f_{beat2}$) are amplified using a current preamplifier (AMP1, As905-1, NF Corporation: frequency bandwidth = DC - 1 MHz, transimpedance gain = $4 \times 10^6$ V/A), followed by an RF amplifier (AMP2, 5307, NF Corporation: frequency bandwidth = DC - 1 MHz).

In the RF-domain signal processing and measurement stage, a portion of the optical output from the dual RI-sensing OFCs is extracted and detected as the repetition frequencies ($f_{rep1}$ and $f_{rep2}$) using a photodetector (PD, PDA05CF2, Thorlabs: wavelength = 800 - 1700 nm, RF bandwidth = DC - 150 MHz). The $f_{rep1}$ and $f_{rep2}$ signals are measured using an RF frequency counter (RF-FC, 53131A, Keysight Technologies: frequency range = 225 MHz) synchronized with the same rubidium (Rb) frequency standard. In parallel, the beat signals ($f_{beat1}$ and $f_{beat2}$) are also measured using the RF-FC. Finally, the differential frequencies are then calculated as $\Delta f_{rep}$ and $\Delta f_{beat}$.

## 4. Results

### 4.1 Verification of frequency multiplication and temperature-drift compensation

In this section, we experimentally verify the frequency-multiplication effect in the THz domain and the effectiveness of active–dummy temperature-drift compensation. In this experiment, both MMI sensors in the active and dummy OFCs were immersed in pure water (refractive index = 1.3180 RIU), ensuring that no RI-induced variation was introduced. Figure 4(a) shows the temporal evolution of the repetition frequencies ($f_{rep1}$ and $f_{rep2}$) and their differential frequency ($\Delta f_{rep}$), measured with a gate time of 100 ms and a sampling interval of 1 s. The red and blue traces represent $f_{rep1}$ and $f_{rep2}$, respectively, while the green trace shows the differential frequency $\Delta f_{rep}$. Even when the fiber cavities are temperature-stabilized, noticeable frequency drift is observed in both $f_{rep1}$ and $f_{rep2}$ on this timescale. This drift limits the sensing performance in single-OFC measurements using $f_{rep1}$ alone. Both $f_{rep1}$ and $f_{rep2}$ exhibit similar fluctuations due to temperature drift, indicating a common-mode response, whereas $\Delta f_{rep}$ remains relatively stable, with a standard deviation of 1.048 Hz, demonstrating effective temperature compensation. However, since the repetition frequencies are directly measured in this case, the frequency-multiplication effect is not yet manifested.

To investigate the frequency-multiplication effect, we next analyze the beat frequencies obtained via photoconductive THz heterodyne detection. Figure 4(b) shows the temporal evolution of the beat frequencies ($f_{beat1}$ and $f_{beat2}$) and their differential frequency ($\Delta f_{beat}$). The red and blue traces represent $f_{beat1}$ and $f_{beat2}$, respectively, while the green trace shows the differential frequency $\Delta f_{beat}$. A behavior similar to that of $f_{rep1}$ and $f_{rep2}$ is observed, where both exhibit fluctuations originating from temperature drift. However, the magnitude of the frequency fluctuations is significantly increased because $f_{beat1}$ and $f_{beat2}$ are generated by heterodyne mixing between the CW-THz signal and the high-order harmonics of $f_{rep1}$ and $f_{rep2}$. The multiplication factor $m$ is defined by the ratio of the CW-THz frequency to the repetition frequencies. In the present experiment, the harmonic of $f_{rep1}$ and $f_{rep2}$ nearest to $f_{THz}$ = 101.270 GHz corresponds to $m$ = 3300. Accordingly, the fluctuation amplitudes of $f_{beat1}$ and $f_{beat2}$ are enhanced by approximately the same factor compared with those of $f_{rep1}$ and $f_{rep2}$. In this way, a 3300-fold frequency-multiplication effect is confirmed. Although the effect is demonstrated here using temperature-induced fluctuations, it is expected to apply equally to RI-induced repetition frequency shifts.

We further evaluate the effectiveness of active–dummy temperature-drift compensation under the frequency-multiplied condition in the THz domain. An important observation in Fig. 4(b) is that the temperature-drift-induced fluctuations in $f_{beat1}$ and $f_{beat2}$, which are amplified by frequency multiplication, remain highly correlated, indicating that their behavior is preserved as a common-mode response. As a result, the differential beat frequency $\Delta f_{beat}$ effectively cancels the temperature-induced fluctuations, similar to $\Delta f_{rep}$. The resulting frequency fluctuation of $\Delta f_{beat}$ has a standard deviation of 3459 Hz, confirming that the differential signal also approximately follows the frequency-multiplication scaling of $\Delta f_{beat} = m\Delta f_{rep}$. These results demonstrate that both frequency multiplication and temperature-drift compensation can be simultaneously achieved in the proposed system.

### 4.2 Performance evaluation of RI sensing

Having verified the frequency-multiplication effect and temperature-drift compensation, we now quantitatively evaluate the RI sensing performance of the proposed system. To quantitatively evaluate the RI sensing performance, sensitivity, linearity, resolution, and accuracy were calculated based on the RI dependence of $f_{rep1}$, $\Delta f_{rep}$, $f_{beat1}$, and $\Delta f_{beat}$ measurements, and used as metrics for comparison. The measurements were performed using ethanol–water solutions with ethanol concentrations of 0%, 2%, 4%, 6%, 8%, and 10%, corresponding to RI values of

1.3180, 1.3187, 1.3193, 1.3200, 1.3206, and 1.3213 RIU, respectively. The RI values were calculated from literature refractive indices of pure water and ethanol based on their mass fractions, and the solutions were prepared by gravimetric mixing using an electronic balance to ensure accurate concentration control. Five repeated measurements were performed for each sample. The sensitivity was determined from the slope of the linear fitting between the sample RI and the averaged frequency shift, while the linearity was evaluated by the coefficient of determination ($R^2$) of the linear fitting. The resolution was defined as the ratio of the average standard deviation of five repeated measurements at each ethanol–water concentration (excluding pure water) to the sensitivity. The accuracy was calculated as the root-mean-square error (RMSE) between the measured data and the fitted values.

We first evaluate the RI sensing performance of a single RI-sensing OFC [see Fig. 1(a)] as a baseline reference. Figure 5(a) shows the RI dependence of $f_{rep1}$. Although a linear trend is observed for the averaged values, a large scatter is present at each concentration due to temperature drift. This is because the frequency fluctuation induced by temperature drift is comparable to or larger than the RI-dependent frequency shift. Consequently, when the measurement time is on a timescale where temperature drift cannot be neglected, its effect is superimposed on the RI-induced signal. The corresponding performance metrics are as follows: the sensitivity is $1.23 \times 10^4$ Hz/RIU, the linearity is $R^2 = 0.8995$, the resolution is $9.82 \times 10^{-4}$ RIU, and the accuracy is $3.95 \times 10^{-4}$ RIU.

We next examine the RI sensing performance of a THz-comb-based frequency-multiplied single RI-sensing OFC [see Fig. 1(b)]. Figure 5(b) shows the RI dependence of $f_{beat1}$. The corresponding performance metrics are as follows: the sensitivity is 4.07 × $10^7$ Hz/RIU, the linearity is $R^2$ = 0.8989, the resolution is 9.87 × $10^{-4}$ RIU, and the accuracy is 3.97 × $10^{-4}$ RIU. Since $f_{beat1}$ corresponds to the $m$-th harmonic component of $f_{rep1}$ ($mf_{rep1}$), the sensitivity is enhanced by a factor of $m$. However, the fluctuations at each concentration are also amplified by the same factor due to frequency multiplication. As a result, no significant improvement is observed in the resolution and accuracy. These results indicate that, under measurement conditions where temperature drift cannot be neglected, this approach alone is not effective for improving overall sensing performance.

We then evaluate the RI sensing performance of a dual RI-sensing OFC employing active–dummy temperature compensation [see Fig. 1(c)]. Figure 5(c) shows the RI dependence of $\Delta f_{rep}$. The corresponding performance metrics are as follows: the sensitivity is 1.60 × $10^4$ Hz/RIU, the linearity is $R^2$ = 0.9970, the resolution is 9.38 × $10^{-5}$ RIU, and the accuracy is 6.12 × $10^{-5}$ RIU. The scatter at each concentration is significantly reduced owing to the active–dummy temperature compensation. As a result, the linearity, resolution, and accuracy are markedly improved compared with those of the single RI-sensing OFC [see Fig. 5(a)]. In contrast, the sensitivity remains comparable to that of the single RI-sensing OFC, as no frequency multiplication is applied, and is therefore much lower than that of the THz-comb-based configuration [see Fig. 5(b)]. Overall, these results demonstrate that the dual-comb approach is highly effective under measurement conditions where temperature drift cannot be neglected; however, the frequency-multiplication effect is not realized in this configuration.

Finally, we evaluate the RI sensing performance of the combined approach, which integrates frequency multiplication with active–dummy temperature compensation in a dual-THz-comb-based configuration [see Fig. 1(d)]. Figure 5(d) shows the RI dependence of $\Delta f_{beat}$. The corresponding performance metrics are as follows: the sensitivity is $5.05 \times 10^{7}$ Hz/RIU, the linearity is $R^2 = 0.9979$, the resolution is $1.07 \times 10^{-4}$ RIU, and the accuracy is $5.50 \times 10^{-5}$ RIU. By comparing with the other panels in Fig. 5, it is evident that both the frequency-multiplication effect and the temperature-compensation effect are successfully realized in the proposed system. For example, compared with Fig. 5(b), a comparable frequency-multiplication effect is achieved, while temperature-induced drift is effectively suppressed by the active–dummy compensation, resulting in a significant improvement in both resolution and accuracy. In contrast, compared with the dual-optical-comb-based RI sensing [see Fig. 5(c)], the linearity, resolution, and accuracy are nearly identical, which may appear to offer limited additional benefit at first glance. However, the key advantage lies in the significantly enhanced sensitivity, which increases the frequency shift from the order of several tens of hertz to several hundreds of kilohertz. This expansion of the measurable frequency shift is crucial for achieving both high-speed and high-precision measurements, as it enables operation beyond the counter limit of RF frequency measurement systems [23].

Table 1 summarizes the comparison of the four approaches in terms of four key performance metrics. This comparison clearly demonstrates that the proposed approach enables simultaneous enhancement of sensitivity, linearity, resolution, and accuracy by combining THz-comb-based frequency multiplication with active–dummy temperature compensation. As a result, it overcomes the conventional trade-off between sensitivity and measurement stability, offering a practical solution for high-speed and high-precision RI sensing.

## 5. Discussion

5.1 Sensitivity scaling and deviation from the ideal multiplication factor

In the present experiment, the frequency multiplication factor is designed to be $m$ = 3300, which is determined by the ratio of $f_{THz}$ to $f_{rep1}$ (or $f_{rep2}$). Based on this value, the sensitivity enhancement factor is expected to be equal to $m$. However, the experimentally obtained enhancement factors show slight deviations from the ideal value: the ratio of the sensitivities of $f_{rep1}$ and $f_{beat1}$ is 3309, while that of $\Delta f_{rep}$ and $\Delta f_{beat}$ is 3156.

The origin of this discrepancy is discussed below. First, the multiplication factor $m$ itself is uniquely determined by $f_{THz}$ and $f_{rep1}$, and is therefore not expected to contain significant uncertainty. Although $f_{rep1}$ exhibits fluctuations on the order of ~100 Hz due to temperature drift [see Fig. 4(a)], this corresponds to only ~330 kHz in the THz domain after multiplication [see Fig. 4(b)]. This value is sufficiently smaller than the THz-comb mode spacing (~ 30 MHz), indicating that the PC-THz comb mode corresponding to the $m$-th harmonic ($mf_{rep1}$) does not cross the CW-THz frequency during the measurement. Therefore, transitions between adjacent harmonic orders ($m \rightarrow m \pm 1$) can be excluded, and the value of $m$ is considered to be well-defined.

The deviation is therefore more reasonably attributed to the uncertainty in the estimation of the sensitivity enhancement factor. In the experiment, the sensitivity is obtained from linear fitting of the frequency shift with respect to the RI. The estimation accuracy of the slope is affected by several factors, including frequency noise, data scatter, and the limited number of measurement points. In particular, the RI dependence of $f_{rep1}$ and $f_{beat1}$ does not exhibit sufficiently high linearity due to relatively large frequency fluctuations, which introduces uncertainty in the fitted sensitivity. Although the differential signals $\Delta f_{rep}$ and $\Delta f_{beat}$ exhibit much higher linearity, the agreement with the ideal multiplication factor $m$ requires an extremely high accuracy in slope estimation, on the order of less than 0.03% (< 1/3300). Even small residual fitting errors can therefore lead to noticeable deviations in the sensitivity ratio. Consequently, the experimentally obtained sensitivity enhancement factors should be interpreted as ratios between sensitivities that inherently include fitting uncertainties.

The observed discrepancy from the ideal value $m$ = 3300 is thus primarily attributed to the limited accuracy in sensitivity estimation, rather than to any intrinsic error in the physical frequency multiplication process itself.

### 5.2 Positioning of this work with respect to previous studies

The present work can be understood as a unification and extension of our two previous approaches: THz-comb-based frequency multiplication [23] and dual-comb active–dummy temperature compensation [22,24].

**Comparison with single-THz-comb-based RI sensing.** In our previous work on THz-comb-based sensing, frequency multiplication enabled a substantial enhancement of sensitivity by scaling the RI-induced frequency shift by a factor of $m$. However, since both the signal and noise were amplified simultaneously, the overall resolution and accuracy were not improved under practical measurement conditions where temperature drift is non-negligible. In contrast, the present work achieves an improvement of approximately one order of magnitude in both resolution and accuracy, while maintaining a comparable level of sensitivity. This improvement stems from the ability of the proposed approach to decouple signal amplification from noise suppression, thereby overcoming the conventional trade-off between sensitivity and stability. Specifically, frequency multiplication in the THz domain enhances the RI-induced frequency shift, whereas the dual-comb active–dummy configuration suppresses temperature-induced fluctuations through common-mode rejection. This orthogonal control of signal scaling and noise suppression enables the simultaneous realization of high sensitivity and high stability, which cannot be achieved by either approach alone.

**Comparison with dual-optical-comb-based RI sensing.** Our previous approach employing a dual-comb configuration effectively suppresses temperature-induced fluctuations through common-mode rejection, resulting in improved linearity, resolution, and accuracy. Nevertheless, the sensitivity remains limited because no frequency multiplication is applied. The present study simultaneously achieves enhanced sensitivity and suppressed noise by integrating frequency multiplication with active–dummy temperature compensation. This can be interpreted as an orthogonal control of signal scaling and noise suppression: frequency multiplication selectively enhances the RI-induced frequency shift, whereas the dual-comb active–dummy configuration suppresses temperature-induced fluctuations. As a result, the proposed system overcomes the conventional trade-off between sensitivity and stability, which cannot be achieved by the dual-comb approach alone.

At first glance, the linearity, resolution, and accuracy appear comparable to those of the dual-comb-based approach, which may obscure the advantage of the proposed method. However, the enhanced sensitivity leads to a substantial increase in the measurable frequency shift, from the order of several tens of hertz to several hundreds of kilohertz. This expansion provides a critical practical advantage, as it enables reliable and rapid frequency readout using RF frequency counters, whose measurement precision is fundamentally limited by the gate time. Consequently, larger frequency shifts allow shorter gate times without sacrificing measurement precision, thereby enabling both high-speed and high-accuracy sensing.

To summarize the above discussion, Table 2 compares the four approaches in terms of sensitivity, resolution/accuracy, and measurement speed. The results clearly indicate that neither single-THz-comb-based frequency multiplication nor dual-comb active-dummy temperature compensation alone is sufficient to simultaneously achieve high sensitivity and high overall measurement performance. In contrast, the proposed dual-THz-comb approach uniquely enables both high sensitivity and high resolution/accuracy, while also significantly improving measurement speed. This capability arises from the orthogonal control of signal scaling and noise suppression, enabled by the integration of THz-comb-based frequency multiplication and dual-comb active–dummy temperature compensation. These results highlight that the proposed method provides a comprehensive solution for high-speed and high-accuracy RI sensing, overcoming the fundamental limitations of the individual approaches.

5.3 Pathways toward further improvement of sensing performance

Although the proposed dual-THz-comb-based approach enables simultaneous high sensitivity, high resolution/accuracy, and high-speed operation, further performance improvement is still possible. In the present implementation, the active and dummy OFCs are not perfectly co-located or fully shared, and therefore residual temperature-induced fluctuations are not completely identical between the two paths. As a result, the common-mode rejection is not perfect as shown in Fig. 5(d), which limits the achievable resolution and accuracy. In the present mechanically shared dual-OFC configuration, common-mode temperature drift can be efficiently suppressed; however, complete cancellation is inherently difficult as long as non-shared components remain in the system. Indeed, slight discrepancies in the temporal behavior of the two OFCs are observed in Fig. 4(b), resulting in residual fluctuations in $\Delta f_{beat}$. These residual fluctuations ultimately limit the achievable sensing performance, indicating an intrinsic limitation of the dual-OFC configuration.

Further improvement can be expected by employing a configuration in which the two OFCs share the same optical cavity. In such a system, temperature fluctuations would be more strongly correlated, leading to more effective cancellation in the differential measurement. One promising approach to overcome these limitations is the use of a single-cavity dual-comb configuration, in which both OFCs are generated within a common cavity [30]. Such architectures are expected to provide significantly improved noise correlation and enhanced stability, enabling further improvement in both resolution and accuracy. For example, residual fluctuations in $\Delta f_{rep}$ on the order of 0.1 Hz have been reported [30], which are approximately one order of magnitude smaller than those observed in the present dual-cavity dual-comb configuration (~1

Hz). While incorporating an MMI fiber sensor directly into a single-cavity configuration is technically challenging, due to issues such as loss, stability, and coherence preservation, practical implementations may be achieved through optimized intracavity integration or hybrid shared-cavity designs. These considerations suggest that the proposed concept of orthogonal control of signal scaling and noise suppression can be further strengthened through improved system integration, establishing a scalable and physically grounded pathway toward next-generation ultra-high-performance RI sensing.

## 6. Conclusion

In this study, we have presented a unified RI sensing platform that integrates THz-comb-based frequency multiplication with dual-comb active–dummy temperature compensation. By combining these two complementary approaches, the proposed system enables simultaneous enhancement of sensitivity and suppression of temperature-induced noise in a single measurement framework.

Experimental results demonstrate that the proposed method achieves a sensitivity on the order of $10^7$ Hz/RIU, while maintaining high linearity ($R^2 \approx 0.998$), improved resolution (~$10^{-4}$ RIU), and high accuracy (~$10^{-5}$ RIU). In addition, the frequency shift is significantly expanded from the order of tens of hertz to hundreds of kilohertz, enabling rapid and precise readout using RF frequency counters with short gate times. These results confirm that the proposed approach successfully overcomes the limitations of conventional RI-sensing OFC systems, in which sensitivity, stability, and measurement speed are inherently coupled. More fundamentally, this work establishes orthogonal control of signal scaling and noise suppression as a design principle for high-performance RI sensing. Frequency multiplication in the THz domain selectively amplifies the RI-induced signal, whereas the dual-comb active–dummy configuration suppresses temperature-induced fluctuations through common-mode rejection. This decoupling of signal amplification and noise suppression enables the simultaneous realization of high sensitivity, high stability, and high-speed operation, which cannot be achieved by either approach alone.

The proposed concept provides a scalable pathway toward next-generation RI sensing systems and is expected to be applicable to a wide range of sensing scenarios, including real-time biochemical reactions, gas monitoring, and dynamic fluid analysis. Furthermore, continued advancements in system integration, such as single-cavity dual-comb architectures, are expected to further enhance performance, opening new possibilities for ultra-high-precision and high-speed optical sensing technologies.

**Funding.** Japan Society for the Promotion of Science (26KJ1756); Japan Science and Technology Agency (JPMJSP2113).

**Disclosures.** The authors declare no conflict of interest.

**Data availability.** Data underlying the results presented in this paper are not publicly available at this time but may be obtained from the authors upon reasonable request.

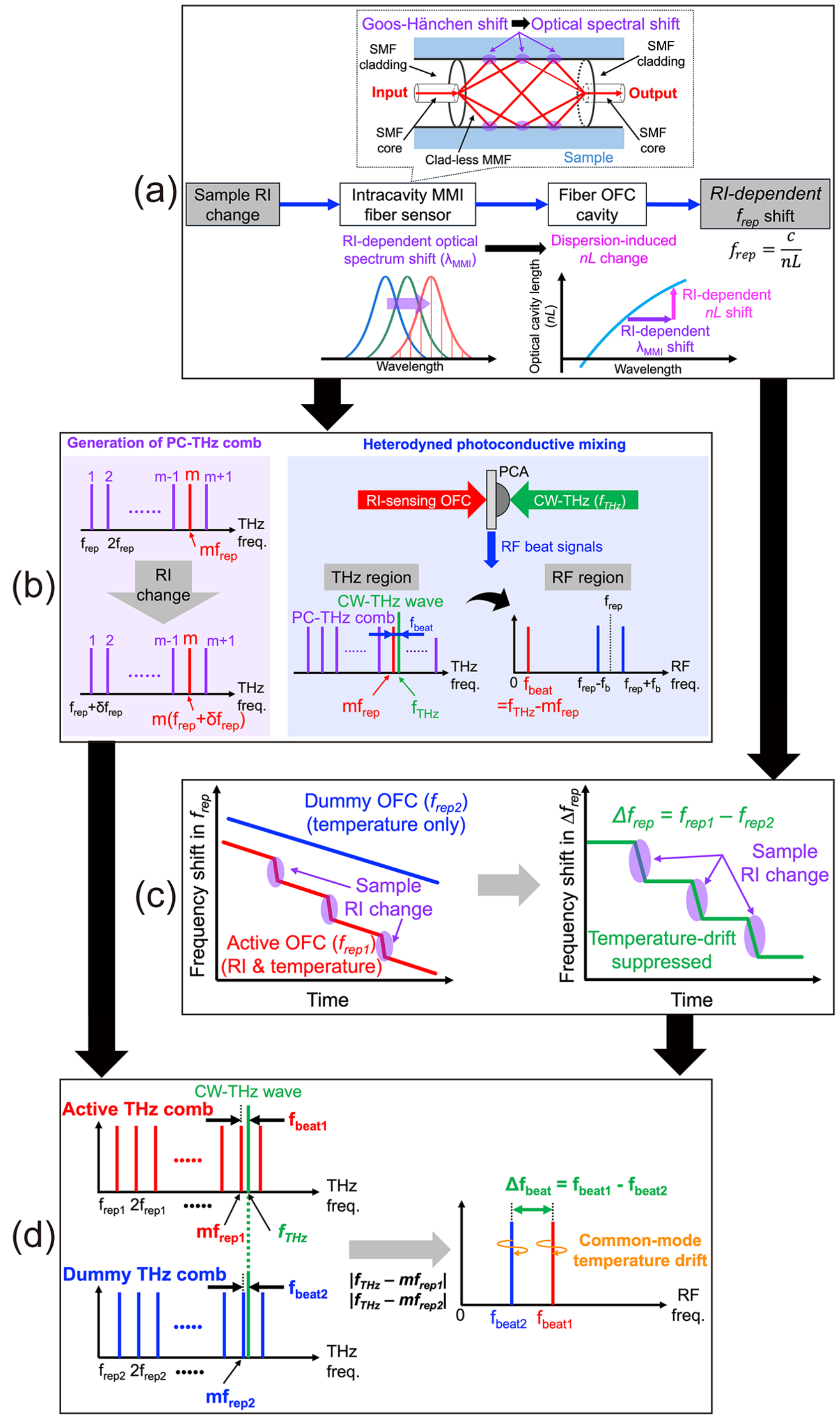

Fig. 1. Operating principles of the proposed RI-sensing system. (a) RI-sensing OFC based on an intracavity MMI fiber sensor, (b) sensitivity enhancement via THz-comb-based frequency multiplication, (c) active–dummy temperature compensation using a dual-OFC configuration, and (d) proposed dual-THz-comb-based approach integrating frequency multiplication with active–dummy temperature compensation.

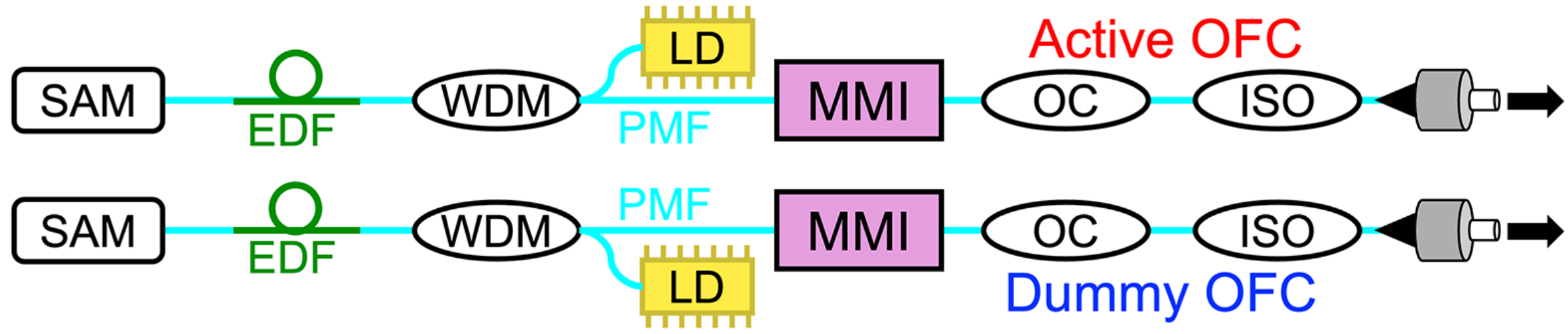

Fig. 2. Configuration of the dual RI-sensing OFC sources. SAM, saturable absorber mirror; EDF, erbium-doped polarization-maintaining single-mode fiber; WDM, wavelength-division multiplexer; LD, laser diode; PMF, polarization-maintaining single-mode fiber; MMI, multimode interference fiber sensor; OC, output coupler; ISO, optical isolator.

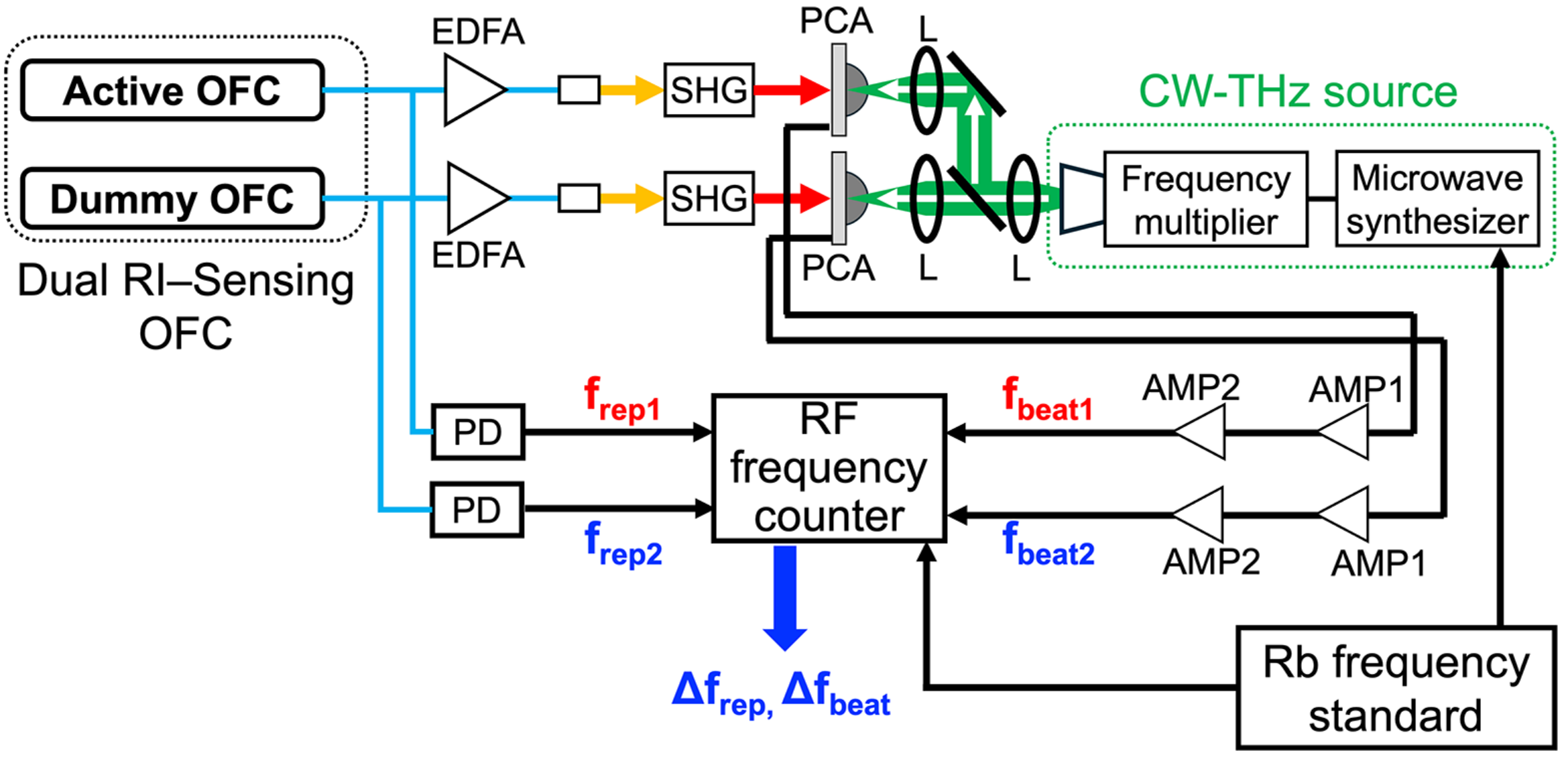

Fig. 3. Experimental setup of the photoconductive THz heterodyne detection system using dual RI-sensing OFCs. EDFA, erbium-doped fiber amplifier; SHG, second-harmonic-generation crystal; PCA, photoconductive antenna; L, Teflon lens; PD, photodetector; AMP1, current preamplifier; AMP2, RF amplifier.

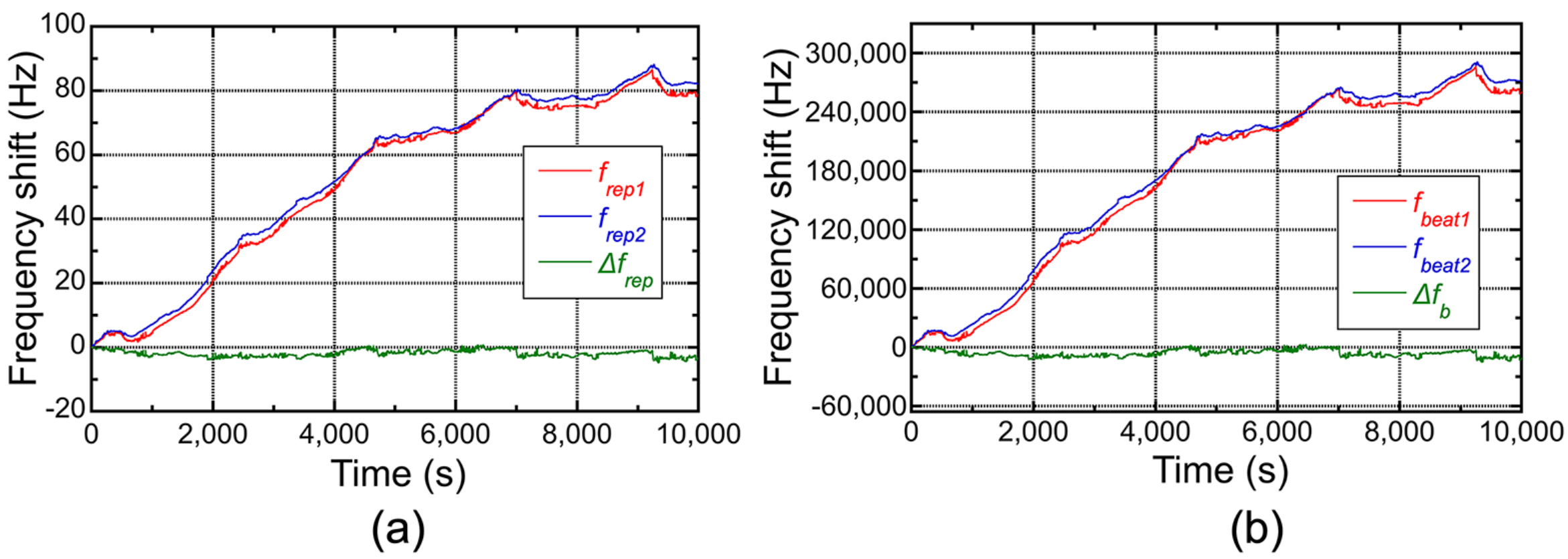

Fig. 4. Experimental verification of frequency multiplication and temperature-drift compensation. (a) Temporal evolution of the repetition frequencies $f_{rep1}$ and $f_{rep2}$, and their differential frequency $\Delta f_{rep}$, measured with a gate time of 100 ms and a sampling interval of 1 s. The red and blue traces represent $f_{rep1}$ and $f_{rep2}$, respectively, while the green trace shows $\Delta f_{rep}$. Common-mode temperature-induced fluctuations are effectively suppressed in the differential signal. (b) Temporal evolution of the beat frequencies $f_{beat1}$ and $f_{beat2}$, and their differential frequency $\Delta f_{beat}$, obtained via photoconductive THz heterodyne detection. The red and blue traces represent $f_{beat1}$ and $f_{beat2}$, respectively, while the green trace shows $\Delta f_{beat}$.

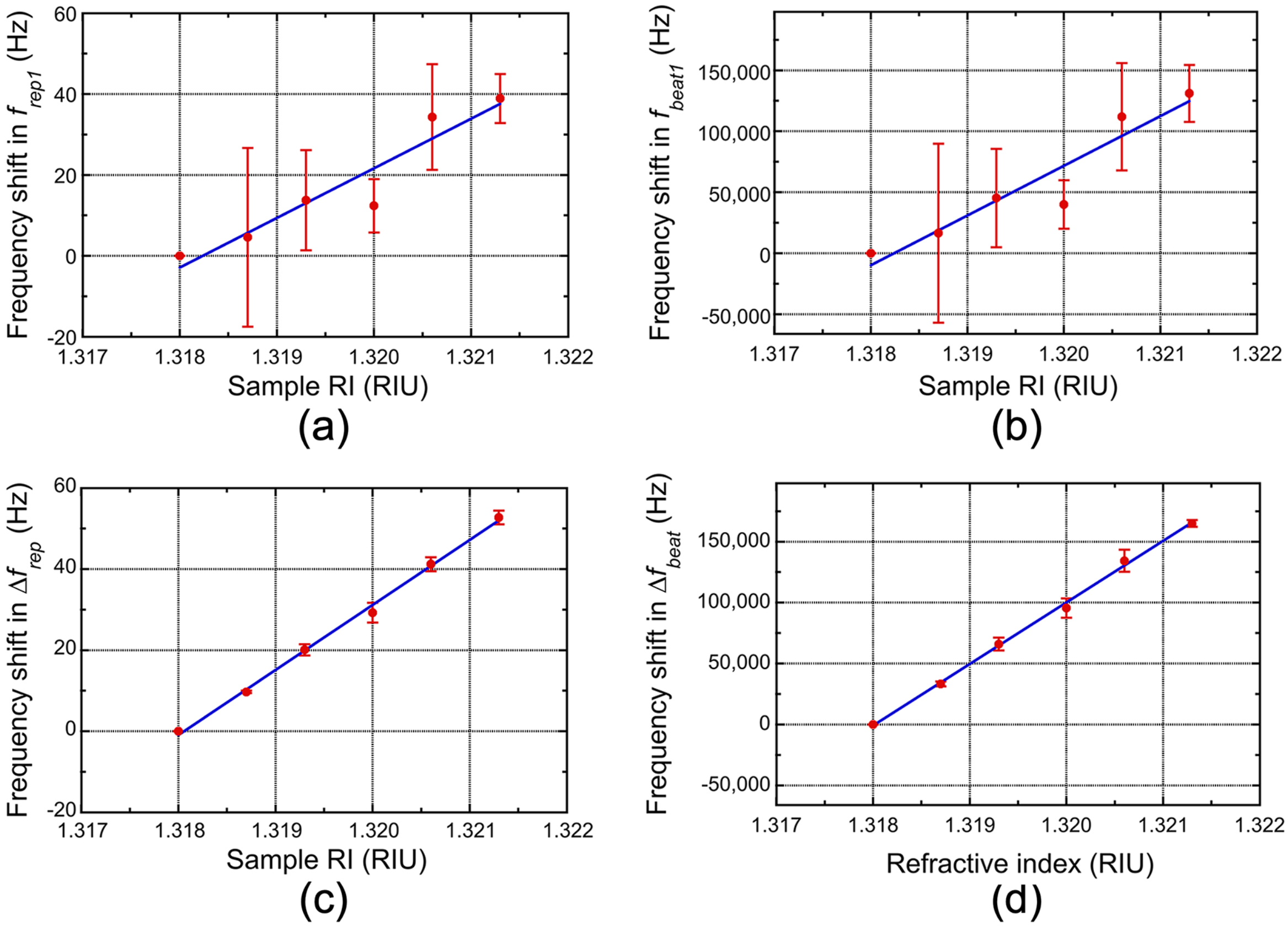

Fig. 5. Comparison of RI sensing performance for different configurations. (a) RI dependence of the repetition frequency $f_{rep1}$ in a single RI-sensing OFC, (b) RI dependence of the beat frequency $f_{beat1}$ in a THz-comb-based frequency-multiplied single RI-sensing OFC, (c) RI dependence of the differential repetition frequency $\Delta f_{rep}$ in a dual RI-sensing OFC with active–dummy temperature compensation, and (d) RI dependence of the differential beat frequency $\Delta f_{beat}$ in the proposed dual-THz-comb-based configuration integrating frequency multiplication and temperature compensation. The blue solid lines represent linear fits, and the error bars indicate the standard deviation obtained from five repeated measurements.

Table 1. Comparison of RI sensing performance among four configurations.

| | Measured quantity | Sensitivity (Hz/RIU) | Linearity | Resolution (RIU) | Accuracy (RIU) |
|---|---|---|---|---|---|
| Single-OFC-based RI sensing | $f_{rep1}$ | $1.23 \times 10^4$ | $R^2 = 0.8995$ | $9.82 \times 10^{-4}$ | $3.95 \times 10^{-4}$ |
| Single-THz-comb-based RI sensing (frequency multiplication) | $f_{beat1}$ | $4.07 \times 10^7$ | $R^2 = 0.8989$ | $9.87 \times 10^{-4}$ | $3.97 \times 10^{-4}$ |
| Dual-OFC-based RI sensing (temperature compensation) | $\Delta f_{rep}$ | $1.60 \times 10^4$ | $R^2 = 0.9970$ | $9.38 \times 10^{-5}$ | $6.12 \times 10^{-5}$ |
| Dual-THz-comb-based RI sensing (frequency multiplication + temperature compensation) | $\Delta f_{beat}$ | $5.05 \times 10^7$ | $R^2 = 0.9979$ | $1.07 \times 10^{-4}$ | $5.50 \times 10^{-5}$ |

Table 2. Qualitative comparison of RI sensing performance among four configurations

| | Sensitivity | Resolution Accuracy | Measurement speed (gate-time limited) | Key characteristics |
|---|---|---|---|---|
| Single-OFC-based RI sensing | Low | Limited (drift-limited) | Low | Simple, but strongly affected by temperature drift |
| Single-THz-comb-based RI sensing | Very high (×m) | Limited (drift-limited) | High | Signal amplified, but SNR unchanged |
| Dual-OFC-based RI sensing | Low | High (noise-suppressed) | Low | Effective noise suppression, but no signal amplification |
| Dual-THz-comb-based RI sensing | Very high (×m) | High (noise-suppressed) | High | Orthogonal control of signal scaling and noise suppression |